# A Framework for the Automated Parameterization of a Sensorless Bearing Fault Detection Pipeline

Tobias Wagner
Dept. of Electrical drives and Controls
Bosch Rexroth AG
Ulm, Germany
ORCID: 0000-0002-4548-6759

Alexander Gepperth
Dept. of Applied Computer Science
Fulda University of Applied Sciences
Fulda, Germany

Elmar Engels
Dept. of Electrical Engineering
Fulda University of Applied Sciences
Fulda, Germany

*Abstract*—**This study proposes a framework for the automated hyperparameter optimization of a bearing fault detection pipeline for permanent magnet synchronous motors (PMSMs) without the need of external sensors.
A automated machine learning (AutoML) pipeline search is performed by means of a genetic optimization to reduce human induced bias due to inappropriate parameterizations. For this purpose, a search space is defined, which includes general methods of signal processing and manipulation as well as methods tailored to the respective task and domain.
The proposed framework is evaluated on the bearing fault detection use case under real world conditions. Considerations on the generalization of the deployed fault detection pipelines are also taken into account. Likewise, attention was paid to experimental studies for evaluations of the robustness of the fault detection pipeline to variations of the motors working condition parameters between the training and test domain.**

The present work contributes to the research of fault detection on rotating machinery in the following terms:
- **Reduction of the human induced bias to the data science process, while still taking into account expert and task related knowledge - ending in a generic search approach**
- **Tackling the bearing fault detection task without the need for external sensors (sensorless)**
- **Learning a domain robust fault detection pipeline applicable to varying motor operating parameters without the need of re-parameterizations or fine-tuning**
- **Investigations on working condition discrepancies with an excessive degree to determine the pipeline limitations regarding the abstraction of the motor parameters and the pipeline hyperparameters**

## I. INTRODUCTION

With more than 50%, bearings are the most frequent cause of failures in electrical machines [1]. Current research pays attention mainly to vibration sensor data analysis, which limits the use of the systems since these sensors usually have to be installed as additional components on the motors. Therefore, in the last decade, research was carried out on using motor internal data for fault detection. Focus hereby was on the phase current data, since this signals are available for most motors ex works. While vibration signals can be efficiently analyzed via frequency analysis, phase current data does not provide such meaningful fault indicators for real damages [2]. Also, phase current signals are stronger subjected to the influence of motor parameters, such as rotational shaft frequency variations, than vibration signals. This hardens the working condition transfer which is of great interest for real world use cases. In order to still perform reliable predictions based on the phase current data, meticulous coordination of all involved hyperparameters, for data pre-processing as well as the classification algorithm, must be ensured.

Therefore, in our research, we propose a fault detection framework with automated hyperparameter selection. For this purpose, task independent methods are bundled with expert driven methods related to the task of the bearing fault detection. In contrast to existing research, our approach considers both, the pipeline performance on unchanged working conditions as well as a working condition transfer performance without the need of re-adjusting any pipeline hyperparameters after the training.

### A. Motivation of phase current based bearing fault detection

Mostly six reasons for damages of bearing components exist. These are a misalignment between shaft and periphery (called eccentricity), overloads to the shaft, resonance vibrations due to loose mounted parts, insufficient greasing, low lubrication quality and overheating. The current state of research mostly considers explicit single-point bearing faults: Defects in the raceways of the inner or outer ring as well as damages on the rolling elements or the cage. For these types of damages, correlations between damage and frequency components, so-called ball pass frequencies (BPFs) exist for vibration as well as current based data [3]. BPFs are sideband frequencies around the harmonics of the drives supply frequency whose amplitudes increase with progressing damage. The authors in [4] made use of the BPFs analyzing the frequency spectrum, using wavelet packet decomposition. Their approach outperformed the performance of Fourier based techniques. Besides the use of MCSA based methods, past research also applied data driven or *learning* techniques to analyse the BPF related fault indicators. In [5] the frequency information was pre-processed by means of a wavelet packet transform. The so extracted information was then forwarded to a 1d convolutional neural network (CNN) for feature extraction and classification. However, the experimental evaluation of the approach is considered to be incomplete since the influence of varying loads is not investigated. However, this causes a covariate shift which in turn can significantly influence the model performances [2]. Existing research on phase-current based bearing fault detection primary focus on BPF based fault indicators. Nevertheless, the presence of these BPFs as fault indicators depends on some pre-conditions which are uncommon in real world applications [6]. Thus, the existence of BPFs is only validated for the mentioned single point fault types. However, in practice most faults belong to the class *general roughness*, which bundles a wide variety of damages of all bearing parts, including multi point damages [2, 7].

*B. Data driven bearing fault detection*

Phase current data is of lower focus in research compared to vibration data. The authors in [8] took into account the phase-current based BPFs for extracting fault indicating features. The approach in [9] used procedures of the so-called motor current signature analysis (MCSA) to classify the energy of individual fault indicators derived from the spectral range. Several deep learning approaches are limited to vibration data only. Thus, in [10], a neural network based on convolution layers was proposed to detect bearing faults of vibration signals corrupted with noise to simulate industrial applications. Random sampling was applied in [11] for more robustness on noise disturbances. To further increase the models robustness on data acquired under rough environments, bundling of multiple sensor sources is applied to achieve more stable prediction performances [12].

A major requirement for fault detection solutions with real world applicability is the ability to abstract variations of the motors working condition parameters like speed and radial forces. In data science nomenclature this is referred to as *covariate shift* and describes the challenge that the feature spaces of the training data (called *source*) differ from the feature representation of the test data (referred to as *target* domain). Related work frequently considers this working condition task by use of deep neural networks. To do this, a supervised classification branch is trained on the source data, while a second branch is fed with the unsupervised target domain data. The overall loss function takes into account both, the loss of the classification stage as well as the so called domain discrepancy, which enforces an alignment of the feature distributions of both domains. The frequency domain features of vibration signals have been enforced to a common feature space in [13] by use of a domain adaptation model. The authors in [14] go one step further and do not only abstract working condition changes but also different motor types. However, the domain discrepancy reduction approaches require a fine-tuning of the models hyperparameters when new working condition data is available.

*C. Contribution and originality*

In related work, some assumptions are made, which are not given in real-world application scenarios. A key pre-requisite of most approaches is that the damages are of single point type without any deviations from the expected norm, like repetitions or multiple damage locations. To fulfill this requirements, the bearings are prepared artificially, e.g. with boreholes. The meaningful BPF features can only be extracted for these types of damages, which in turn do not match the damages that occur in real-world applications [2]. Another common assumption relates to the working condition transfer strategy, taking into account the data of both domains to reduce the feature discrepancies. For this, the label spaces of both domains must be equal. This case is unlikely to occur in industrial use cases since it means that a motor would break down exactly during the data acquisition [15].

Our research proposes a fault detection framework to overcome some of the aforementioned drawbacks by generalizing the prediction task. The idea of the framework is to make use of general signal processing and manipulation knowledge as well as existing domain expertise. Hereby, a toolbox like fundus of data operations and transformations is created which is chained to a full stack data science pipeline, represented as a directed acyclic graph (DAG). The pipeline includes the data pre-processing, feature extraction and optimization as well as the classification. To avoid possible biases like the aforementioned BPF dependency, the selection of the pipelines transformation steps as well as their hyperparameter selection, is done in an automated manner by means of a genetic optimization.

## II. THE PROPOSED FAULT DETECTION FRAMEWORK

The selection of the model hyperparameters to reach an optimal performance is the main goal of each data driven approach. For this purpose, the so called AutoML fundamentals are utilised in the scope of the proposed fault detection framework. By this, the bias due to human-induced mis-parameterization should be reduced by means of a automated parameterization. The remaining human induced bias is limited to the definition of the initial search space which is optimized during a genetic optimization.

*A. Encapsulating the data science workflow*

For condition based maintenance (CBM) systems, the application of automated hyperparameter optimization is currently still rarely used. However, some research yielded promising results and outperformed the baseline performance of models with a a priori hyperparameter definition [16, 17]. However, unlike existing approaches, the proposed framework optimizes the hyperparameter of all operations used in the data science stack instead of restricting hyperparameter selection to the classification level only.

The framework uses chained methods, bundled in groups, to create the fault detection pipeline (DAG). Figure 1 shows the five groups and some of the methods, hereinafter referred to as *transformers*, assigned to them:

1) **Data source preprocessing:** Manipulations on the raw data based on domain expert driven transformers.
2) **General preprocessing:** Performs general transformation on the time series signals like filtering and data augmentation.

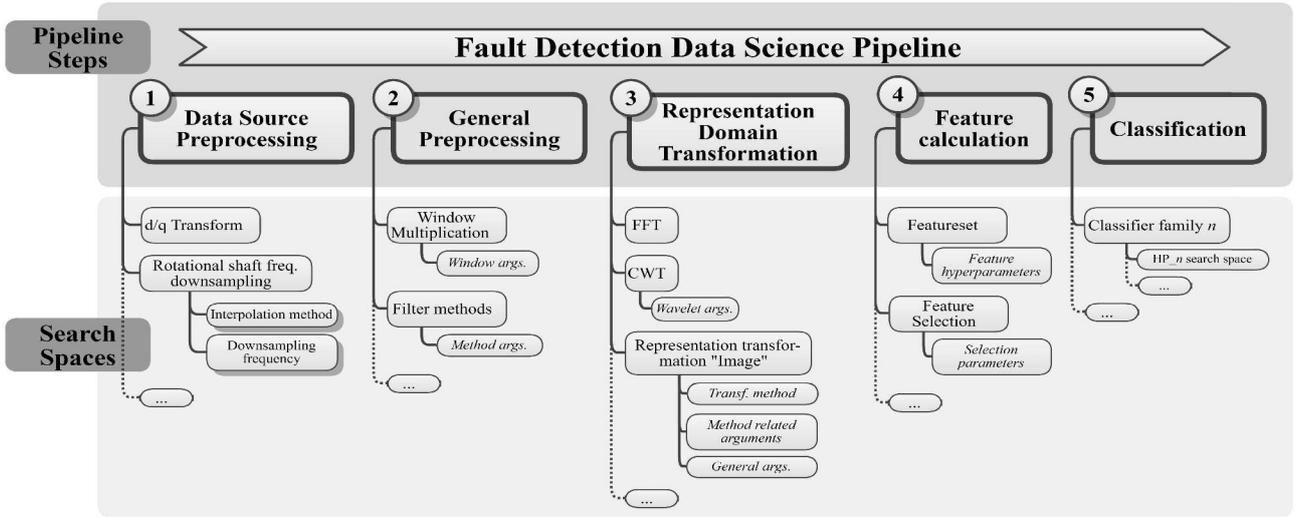

Fig. 1. Proposed fault detection pipeline as a modular tool box for a directed acyclic execution

3) **Representation domain preprocessing:** Transforms the time series signal presentation to alternative representation domains.
4) **Feature calculation:** Extracts a set of statistical features and selects a optimized subset.
5) **Classification:** Selects a hyperparameter optimized classification algorithm to solve the prediction task.

*B. Pipeline transformers*

One advantage of the automated transformer selection is its task-independent and generic structure. This means that methods from past projects can be re-used and evaluated for their suitability in the current task. This leads to a steadily growing bundling of expert-driven approaches for the respective task, as well as general procedures. Selected transformers are introduced in the following.

*1) Data source preprocessing:* A task-specific method, which was developed for the concrete challenge of the working condition transfer is the so called *rotational shaft frequency resampling (RSFR)*. It aims to reduce the discrepancy between samples of different rotational speeds by downsampling to a (fictitious) pseudo rotational speed. The pseudocode of the RSFR is shown by algorithm 1. The term $n_{osc}^{S/T}$ specifies the number of oscillations to select from the original signal (source/target) to form the support points for the interpolation, to reach the number of oscillations according to the requested pseudo speed.

*2) General preprocessing:* Due to the often small number of available samples, the fault detection pipeline also includes a data augmentation stage by chunking the full length samples $X$ in $N_s$ samples of length $s$ according to eq. 1.

$$N_s = \left\lfloor \frac{len(X) - w}{s} + 1 \right\rfloor \quad (1)$$

To further increase the extend of augmentation, a overlapping $w$ can be applied for the training samples as well. For image-like data representations, augmentation strategies like rotation, flipping and cropping/zooming are applied.

**Algorithm 1 RSFR**

**Require:** $X^{S \times O}$, Input: $S$ samples with $O$ observations each
1: $M \leftarrow X$
2: Calculate $f_{ps} = \frac{f_r \cdot n_{polpairs}}{2 \cdot \pi}$
3: Calculate $n_{osc} = \frac{1}{f_s} \cdot len(X) \cdot f_{ps}$
4: Decompose $X$ into interpolation support vectors
5: **for each** *interp_vector* **do**
6:     Interpolate to $n_{osc} = \frac{len(X) \cdot n_{osc}^T}{n_{osc}^S}$ observations
7:     Update $M$ with the new interpolation vector
8: **end for**
**Ensure:** $M$, Transformed $S^*$ samples with $O$ observations

*3) Representation domain transform:* Transforming the raw time series signals to additional representation domains like frequency or time-frequency domain, can improve the selection of reliable fault indicators [18, 19]. The proposed framework therefore transforms the raw signals to frequently used representation domains, like the Fourier spectrum (FFT), the power spectral density (PSD) and the Wavelet domain.
In addition, image representations are also created from the time series data. In total three methods are applied for image translation: Recurrence plots [20] as well as Gramian angular fields and Markov transition fields [21].

*4) Feature calculation:* The proposed framework relies on a feature extraction strategy based on statistical characteristics. The full size feature set $\mathcal{X}^n = \{x_1, \ldots, x_n\}$ bundles the $n = 24$ initial features $x \in \mathbb{R}$ which were found to be suitable for fault detection on rotating machinery in related work [22,

7, 23]. An optional dimensionality reduction by selecting the most significant features can be performed within the pipeline optimization. For this, the results of multiple feature selection strategies like *PCA* [24] and *mRMR* [25] are compared. Besides the feature selection, further pre-processing steps, including scaling and low-variance cleaning, are performed during the feature set optimization to increase the chances of success of the subsequent classification.

*5) Classification algorithms:* Within the scope of the classification stage, the pipeline optimization aims to approximate a prediction $\widehat{F}(x)$ to achieve a minimal expected error rate $\mathbb{E}_{x,y}$ on unseen samples. The proposed framework performs the classifier hyperparameter search within the pipeline optimization, which is not the case in related research [16, 17]. For the experimental results, mostly the gradient boosting (GB) algorithm was used. We selected this type of algorithm since it was applied in related work on bearing fault detection tasks with promising results [26]. Nevertheless, other algorithms can also be integrated in the pipeline, as it is shown in the later sections.

### C. Genetic hyperparameter optimization

The fault detection pipeline is represented as a acyclic directed graph (DAG) of the aforementioned transformers. Each transformer has at least one hyperparameter, namely for its activation. In terms of *genetic optimization*, the set of all possible solutions is represented as search space $\mathcal{X}$, while each individual solution $x$, referred to as *chromosome*, is composed by $n$ genes $(g)$: $x = \{g_1, g_2, \ldots, g_n\} \subseteq \mathcal{X}$. The fitness function $f$ to be minimized, evaluates the suitability of each solution $V = min(f(g_1, g_2, \ldots, g_n))$. The genetic optimization was already successfully applied in related work on rotating machinery fault diagnosis [27]. For detailed information on the genetic optimization itself, we refer to the further literature [28, 29].

A solution candidate $s_i$ is a chain of transformers (represented as genes) is shown in eq. 2 for an exemplary pipeline using a *SVM* classifier. The genes $g_{28,29}$ represent the regularization parameter $C$ and the *kernel* type.

$$s_i = [\underbrace{g_1}_{Sensor\ Transform}, \underbrace{g_2}_{Augmentation}, \underbrace{g_3}_{Freq.\ Domain}, \underbrace{g_4, g_6, \ldots, g_{27}}_{Feat.\ Vectors}, \underbrace{g_{28}, g_{29}}_{SVM}] \quad (2)$$

Solution candidates can be of varying complexity. Therefore the applied genetic optimization works in a *multi-objective* way by taking into account the aforementioned *fitness objective* as well as the *pipeline cost*[1]. The optimal solution is *searched* by use of the *non dominated sorting genetic algorithm (NSGA-II)* [30].

### D. Solution selection strategy

In order to select the final solution candidate for pipeline deployment, the optimizations are performed in a 5 fold cross validation (CV) with 3 repeats. The folds are grouped by the individual serial numbers of the used examinee motor instances in non-overlapping manner. The goal of the framework is to abstract the pipeline hyperparameters from the motor instances as well as their operational parameters. In order to ensure this, testing takes furthermore place on an independent holdout set. For investigations on the pipeline robustness on working condition induced covariate shifts, this holdout set is also evaluated on a different (target) working condition. Its worth mentioning that the proposed evaluation strategy differs from related works in terms of transfer studies. Pipelines are selected solely based on the CV results. Target data is only used for the transfer performance evaluation to provide an indication of the expected pipeline performance on unknown working conditions.

[1]We refer with *costs* to the duration of the optimization

## III. EVALUATION ON A REAL WORLD USE CASE

In order to prove that the framework is suitable for data-driven bearing fault detection, experimental studies are presented in this chapter. The goal is to validate the contributions of the research mentioned in the introduction as well as to prove how our work can improve missings in the current state of research.

### A. Dataset

The data acquisition of the examinee motors was performed on a test rig, which was developed with focus on real-world application conditions: different radial forces were applied to the motors at different speeds to simulate a belt operation. The radial forces were applied to the motors shaft end by a pneumatic cylinder. Each combination of speed/radial force is further referred to as working condition. For the transfer robustness, data of 4 working conditions were acquired. The dataset contains data of 22 PMSMs which had a proven bearing damage due to long-term operation. The damages (in literature referred to as *general roughness*) ranged from damages at the treats and rolling elements, to broken cages and other parts. The running life-time of the motors ranged from 3 to 12 years (with continuous maintenance). The examinees were of different motor series with different bearing sizes and types, to evaluate the generalization of the fault detection pipelines.

### B. Baseline results & motivation

Due to the wide range of application scenarios in which PMSMs are used, a major requirement is the robustness of the fault detection solution to working condition variations. Table I shows the intrinsic transfer performance of a pipeline without any pre-processing steps for optimizing the transfer robustness applied. The raw phase current data was augmented by splitting each sample to multiple windows of 1024 observations each without overlapping. Both, the spectral domain data calculated using the FFT, as well as the time series windows, were used to extract the 24 features each. The features were normalized to 0...+1 and classified using a *gradient boosting* classifier.

TABLE I
BASELINE AND INTRINSIC PIPELINE TRANSFER PERFORMANCE

| | Source WC | | Target WC | | Baseline Accuracies [%] | |
|---|---|---|---|---|---|---|
| | Speed [rpm] | Force [N] | Speed [rpm] | Force [N] | Source | Target |
| 1 | 250 | 0 | 2000 | 1000 | 73.91 → | 59.20 |
| 2 | 250 | 1000 | 2000 | 0 | 77.17 → | 62.00 |
| 3 | 2000 | 0 | 250 | 1000 | 75.43 → | 65.80 |
| 4 | 2000 | 1000 | 250 | 0 | 72.17 → | 53.50 |

TABLE II
SEARCH SPACE OF THE PIPELINE OPTIMIZATION

| Stage | Transformer (method) | Hyperparameter (description: value range) | Stage | Transformer (method) | Hyperparameter (description: value range) |
|---|---|---|---|---|---|
| A | A1 (notch filter) | $a_{10}$ (filter frequency): {16.6, 133.3} | C | C1 (spectral represent. selection) | $c_{10}$ (selected method): {FFT, PSD} |
| | A2 (RSFR) | $a_{20}$ (pseudo shaft frequency): {118.0} | | C2 (feature calculation) | initial 24 features acc. II-B4 |
| | A3 (Park transformation) | - | | C3 (feature cleaning by low variance) | - |
| | A4 (selection of A1-A3) | $a_{40}$ (selected method): {Raw, Notch, RSFR, Park} | | C4 (feature scaling) | $c_{40}$ (scaling method): {0,...,+1, Standardized, ZScore} |
| | A5 (Savitky-Golay filter) | $a_{50}$ (window length): {5}; $a_{51}$ (polynomial order): {2, 3} | | | |
| B | B1 (data augmentation) | $b_{10}$ (window size): {1024, 2048}; $b_{11}$ (overlapping): {0} | D | D1 (gradient boosting algorithm) | $d_{10}$ (estimators): {100}; $d_{11}$ (nodes per decision tree): {1, 2, ..., 10}; $d_{12}$ (SGD lr): {$e^{-3}, e^{-2}, e^{-1}, 0.5, 1.0$}; $d_{13}$ (min. weight of a DT node): {1, 2, ..., 20} |
| | B2 (normalization) | - | | | |
| | B3 (detrend) | - | | | |
| | B4 (analytical signal) | - | | | |
| | B5 (window multiplication) | - | | | |

The source baseline results advocate a general feasibility of the phase current based bearing fault detection of the used dataset. However, the accuracies are significantly worse compared to the results of related work [7]. Applying the *source* pipelines on data from a different working condition, resembles random guessed predictions due to the covariate shift between the feature spaces of both working condition domains. The results in table I are referred to as baseline results, which are expected to be outperformed by applying the proposed framework.

### C. Search space preparation

To reduce the influence of human intervention during the solution generation, the proposed framework automates the parameterization of the fault detection pipeline. Nevertheless, a search space must be defined. In order to limit the required system resources as well as to reduce the duration of the optimization, some restrictions of the search space were made in advance. Table II shows the transformers included in the pipeline optimization. For better understanding, the pipeline transformers were divided in the four groups $A - D$. The resulting DAG is shown in figure 2.

A data sample $x_i$ passes the pipeline and is finally classified by the gradient boosting algorithms objective function $f(x_i)$. The dashed lines in fig. 2 represent a omittable transformer step. The genetic optimization was performed with 100 generations and a population size of 10 solutions each.

### D. Results of the optimized pipelines

The results of the pipeline optimization are shown in table III. For a valid comparison with the baseline results, we applied the same setting (train on *source*, test on *target* data). The optimized pipelines outperformed the baseline results for all settings. The highest improvement for the *source* pipeline was reached in setting ④ with an improvement by 13.6 %. This setting and pipeline also reached the highest *target* transfer improvement by 34.3 %. During the optimization, the pipeline results were continuously improved: the worst *source* CV accuracies for the settings ①-④ were: 55 %, 63 %, 52.4 % and 50 %. The wide range between the results during the optimization advocates the proposed automatic optimization strategy. Reaching similar results with *a priori* defined hyperparameters is unlikely due to the large number of combinations: only 8 of all (>100) evaluated DAGs passed an acceptance criteria of 85 %. In contrast to related research, *target* data was only used for testing. The process of the pipeline selection was based on the *source* CV accuracy. The target transfer results in tab. III outperformed the respective intrinsic transfer results from tab. I in all settings. The highest improvement of about 34 % was achieved on setting ④.

TABLE III
RESULTS OF THE OPTIMIZED PIPELINE

| | Source WC | | Target WC | | Accuracy [%] | | Accuracy |
| | Speed [rpm] | Force [N] | Speed [rpm] | Force [N] | Source | Target | impact [%] |
|---|---|---|---|---|---|---|---|
| 1 | 250 | 0 | 2000 | 1000 | 86.72 → | 87.80 | +1.08 |
| 2 | 250 | 1000 | 2000 | 0 | 87.78 → | 77.20 | −10.58 |
| 3 | 2000 | 0 | 250 | 1000 | 87.63 → | 83.30 | −4.33 |
| 4 | 2000 | 1000 | 250 | 0 | 85.83 → | 87.80 | +1.97 |

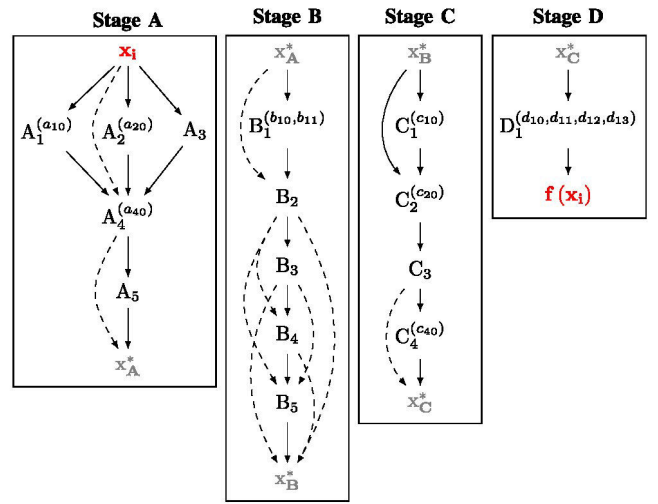

Fig. 2. DAG of the fault detection pipeline, ordered by stages A-D

*E. Impact of an adjusted search space*

Since the search space was restricted in advance, due to resource limitations, we considered the question if these restrictions excluded transformers that would have significantly improved the pipeline performance. Likewise, however, we also considered whether the search space could have been made more efficient without significantly degrading the pipeline performance.

Therefore, we considered adjustments of the following two stages:

- **Stage C - Feature space**: dimensionality reduction
- **Stage D - Classification**: algorithm selection

For the subsequent investigations, the pre-processing pipeline from fig. 2 was applied. Only the respective stage under consideration (**C** or **D**) was adjusted.

*1) Investigations on the feature space:* The initial feature space is spanned by the 24 feature, found to be suitable for bearing fault detection in related research (see. II-B4). We compared four feature selection methods: *Principal component analysis (PCA)* [24, 31], *sequential feature selection (SFS)* [32], *univariate feature selection (UFS)* and *maximum relevance minimal redundance (mRMR)* [25]. The considered methods were already applied by related research for similar bearing fault detection tasks. The feature selection was performed within a range of a minimum of 2 and a maximum of 24 selectable features. Table IV shows the results for the 4 working condition settings. The maximum differences between the four methods are small and lay between 0.78 % and 5.20 %. As a drawback of the feature selection, deterioration's on the pipelines transfer performance were recognized. Compared to the transfer results from table III, the deterioration's were (from setting ① to ④): 15 %, 15 %, 4 % and 7 %. Summarizing, the results showed that feature selection can optimize the pipelines regarding their *source* performance. Considering transfer performances, the results indicate that a sufficient number of features is required to overcome the influence of the remaining domain discrepancies between the feature distributions.

*2) Comparison of classification algorithms:* Due to its promising results in related work, we only applied the GB algorithm in the results from table III. Including multiple classification algorithms within the scope of the pipeline search, significantly increases the overall time required for the optimization. Therefore, we compared the following six algorithms to verify if extensions on the classifier search spaces, could improve the results, compared to the pre-defined GB setting:

- **Ensembles:** Random forest (RF), GB, Extra trees classifier (ETC)
- **Linear models:** Logistic regression (log. regr.)
- **Non parametric models:** k-nearest neighbors (k-nN)
- **Neural network:** Multilayer perceptron (MLP)

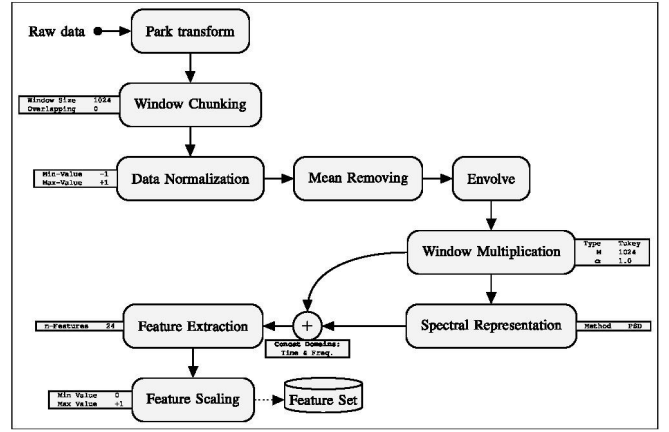

Fig. 3. Pre-defined data transformation pipeline for the classifier comparison

To reduce biases on the results, caused by other transformers than the classification algorithms, a fix pre-processing pipeline according to figure 3 was applied. The genetic optimization of the classifiers was carried out with 100 generations with a population size of 10 solutions each. In total approx. 1000 hyperparameter constellations per classifier were evaluated. Table V shows the results on the four working conditions. Summarizing the results, the differences between the considered classifiers (avg. 7 %) is low. The classifier performances differentiate much more if considering the transfer settings. The GB and Log. Regr. algorithm gained the most stable results over all working condition settings ① to ④. Nevertheless, the results of the Log. Regr. were slightly worse compared to those of the GB. Regarding the results from table III, which were created by use of the GB algorithm, considerations on several classification algorithms gained no improvements. Taking into account the influence of the extended classification stage on the overall optimization duration, therefore indicates that the selection of the classification algorithm does not offer great potential for significant improvements on the overall pipeline performances.

TABLE IV
IMPACT OF FEATURE SELECTION ON THE PIPELINE PERFORMANCES

| Working Condition | | #Features | Selection Method | CV Accuracy[%] | Max. Accuracy Difference [%] |
|---|---|---|---|---|---|
| Speed [rpm] | Radial Force [N] | | | | |
| 250 | 0 | 7 | SFS | 90.66 | 3.58 |
| | 1000 | 12 | SFS | 91.90 | 5.20 |
| 2000 | 0 | 17 | PCA | 91.82 | 1.72 |
| | 1000 | 9 | SFS | 91.06 | 0.78 |

TABLE V
CLASSIFIER COMPARISON ON SOURCE DATA

| Working Conditions | | Classification algorithms Holdout accuracies [%] | | | | | | Max. diff. [%] |
|---|---|---|---|---|---|---|---|---|
| Speed [rpm] | Radial Force [N] | RFC | ETC | GBC | Log. Regr. | k-nN | MLP | |
| 250 | 0 | 83.30 | 83.00 | 90.60 | 89.25 | 88.60 | 84.15 | 7.6 |
| | 1000 | 91.50 | 89.30 | 92.30 | 89.70 | 93.40 | 90.65 | 4.1 |
| 2000 | 0 | 81.90 | 84.55 | 90.50 | 90.60 | 89.60 | 82.85 | 8.7 |
| | 1000 | 86.50 | 80.10 | 82.60 | 87.10 | 87.95 | 83.25 | 7.85 |

## IV. CONCLUSION

The study on our proposed framework mainly contributes to the state of research as follows:

1) Automated creation of the fault detection pipeline to reduce human induced biases
2) The need for external sensors becomes obsolete due to data representation transformations and data manipulations on internal signals
3) Gaining domain robust pipelines by considerations on the source domain only
4) Evaluation of the methodology on the basis of real-world data related to the considered application scenarios

Contrary to related work, the proposed framework makes use of a automated concatenation of several data transformations to create a fault detection pipeline. The results verified, that the pipelines created in automated manner, outperformed the results of handcrafted pipelines. Due to the high number of hyperparameters, our approach thus can assist in making more optimal decisions while creating a fault detection solution. To ensure the practicability of the framework, evaluations were carried out using real-world data instead of artificially prepared bearings. Only motor internal phase-current signals were considered, to overcome the drawbacks of external vibration sensors. Investigations on the transfer performance of the pipelines were carried out using multiple working conditions, which exceeds the scope of comparable research. Hereby, the target domain data was used for evaluation purposes only. This differentiates the presented approach from related research because no re-parameterizations are required if the motors working condition varies between training and inference.